\documentclass{article}
\usepackage{cite}
\usepackage{graphicx}
\usepackage{dcolumn}
\usepackage{amsmath}

\begin{document}

\date{}
\title{On the problem of the vanishing discriminant}
\author{Francisco M. Fern\'{a}ndez\thanks{fernande@quimica.unlp.edu.ar} \\INIFTA, DQT, Sucursal 4, C. C. 16, \\
1900 La Plata, Argentina}
\maketitle

\begin{abstract}
We show that the straightforward application of the discriminant to some
physical problems may yield a trivial useless result. If the symmetry of the
model matrix does not change with variations of the model parameter the
discriminant may vanish for all values of the parameter due to degeneracy.
We illustrate this problem by means of a simple $6\times6$ matrix
representation of an Hermitian Hamiltonian operator.
\end{abstract}

\section{Introduction}

\label{sec:intro}

The resultant of two polynomials and the discriminant of a polynomial are
known since long ago in the mathematical literature\cite{S1851,G81,BPR03}
and have already been applied to the analysis of several physical problems.
Some examples are the determination of singularities in the eigenvalues of
parameter-dependent matrix eigenvalue problems\cite{HS91}, level degeneracy
in a quantum two-spin model\cite{SM98}, the analysis of the properties of
two-dimensional magnetic traps for laser-cooled atoms\cite{D02}, the
description of optical polarization singularities\cite{F04}, the exceptional
points (EPs) for the eigenvalues of a modified Lipkin model\cite{HSG05}, the
location of crossings and avoided crossings between eigenvalues of
parameter-dependent symmetric matrices\cite{BR06,B07,BR07a,BR07b} and the
solution of two equations with two unknowns that appear in the study of
gravitational lenses\cite{TB16}.

The purpose of this paper is to show that these remarkable mathematical
tools are not foolproof and, consequently, should be applied with care. In
section~\ref{sec:example} we introduce a simple parameter-dependent
symmetric matrix and show that the discriminant of its characteristic
polynomial vanishes for all values of the model parameter. We reveal the
reason for the apparent failure of the discriminant and show how to overcome
this difficulty in order to obtain the crossings and avoided crossings
between eigenvalues. In section~\ref{sec:symmetry} we discuss the problem
from the point of view of symmetry and, finally, in section~\ref
{sec:conclusions} we summarize the main results and draw conclusions.

\section{Simple example}

\label{sec:example}

Present discussion is based on the simple $6\times 6$ symmetric matrix
\begin{equation}
\mathbf{H}(\lambda )=\left(
\begin{array}{llllll}
0 & 1 & 0 & 0 & 0 & \lambda \\
1 & 0 & \lambda & 0 & 0 & 0 \\
0 & \lambda & 0 & 1 & 0 & 0 \\
0 & 0 & 1 & 0 & \lambda & 0 \\
0 & 0 & 0 & \lambda & 0 & 1 \\
\lambda & 0 & 0 & 0 & 1 & 0
\end{array}
\right) ,  \label{eq:H}
\end{equation}
that depends on a real parameter $\lambda $. This model was chosen
some time ago for the application of perturbation theory to the
resonance energy given by the H\"{u}ckel model for benzene in
exercise 6.6, page 347, of Szabo and
Ostlund\cite{SO96}. Its real eigenvalues $E_{j}(\lambda )$, $j=1,2,\ldots ,6$%
, are roots of the characteristic polynomial
\begin{equation}
p(E,\lambda )=\left| \mathbf{H}-E\mathbf{I}\right| =E^{6}-3E^{4}\left(
\lambda ^{2}+1\right) +3E^{2}\left( \lambda ^{4}+\lambda ^{2}+1\right)
-\lambda ^{6}-2\lambda ^{3}-1,  \label{eq:charpoly}
\end{equation}
where $\mathbf{I}$ is the $6\times 6$ identity matrix.

The values of $\lambda $ corresponding to level crossings or avoided
crossings can be obtained from the discriminant of the characteristic
polynomial $Disc_{E}\left( p(E,\lambda )\right) $\cite{BR06,B07,BR07a,BR07b}
(see Appendix~\ref{sec:appendix} for definition, notation and properties of
the discriminant). According to equation (\ref{eq:Disc_x(A)}) of the
Appendix~\ref{sec:appendix} this discriminant should enable us to obtain the
values of $\lambda $ for which two (or more) eigenvalues of $\mathbf{H}$
cross. However, in the present case this strategy simply produces the
useless result $Disc_{E}\left( p(E,\lambda )\right) =0$ for all $\lambda $.
In order to understand the reason for this failure note that the
characteristic polynomial can be factorized as
\begin{equation}
p(E,\lambda )=\left( E+\lambda +1\right) \left( E-\lambda -1\right) \left(
E^{2}-\lambda ^{2}+\lambda -1\right) ^{2}.  \label{eq:p_factorized}
\end{equation}
We arbitrarily organize the eigenvalues as
\begin{equation}
E_{1}=-E_{6}=-1-\lambda ,\;E_{2}=E_{3}=-E_{4}=-E_{5}=-\sqrt{\lambda
^{2}-\lambda +1}.  \label{eq:E_j}
\end{equation}
The discriminant of the characteristic polynomial vanishes because $%
E_{2}(\lambda )=E_{3}(\lambda )$ and $E_{4}(\lambda )=E_{5}(\lambda )$ for
all $\lambda $.

We can obtain the desired crossings or avoided crossings from the
discriminant of the polynomial
\begin{eqnarray}
q(E,\lambda ) &=&\left( E+\lambda +1\right) \left( E-\lambda -1\right)
\left( E^{2}-\lambda ^{2}+\lambda -1\right)   \nonumber \\
&=&E^{4}-E^{2}\left( 2\lambda ^{2}+\lambda +2\right) +\left( \lambda
+1\right) ^{2}\left( \lambda ^{2}-\lambda +1\right) ,  \label{eq:q(E,lambda)}
\end{eqnarray}
where we have removed the two-fold degeneracy that is independent of $%
\lambda $. It follows from
\begin{equation}
Disc_{E}\left( q(E,\lambda )\right) =1296\lambda ^{4}\left( \lambda
+1\right) ^{2}\left( \lambda ^{2}-\lambda +1\right) ,  \label{eq:Disc_E(q)}
\end{equation}
that there are actual level crossings at $\lambda =-1,0$ and an avoided
crossing due to the coalescence of eigenvalues in the complex $\lambda $%
-plane at $\lambda =\lambda _{EP}^{\pm }=\left( 1\pm \sqrt{3}i\right) /2$.
The multiple level crossing at $\lambda =0$ comes from the fact that the $%
6\times 6$ matrix exhibits a block diagonal form with three
$2\times 2$ sub matrices of the form
\begin{equation}
\mathbf{H}_{j}=\left(
\begin{array}{ll}
0 & 1 \\
1 & 0
\end{array}
\right) ,\;j=1,2,3,  \label{eq:H_j_block}
\end{equation}
each one with eigenvalues $E=\pm 1$. They are representations of
the three
ethylene molecules in the H\"{u}ckel model discussed by Szabo and Ostlund%
\cite{SO96}.

\begin{figure}[tbp]
\begin{center}
\includegraphics[width=9cm]{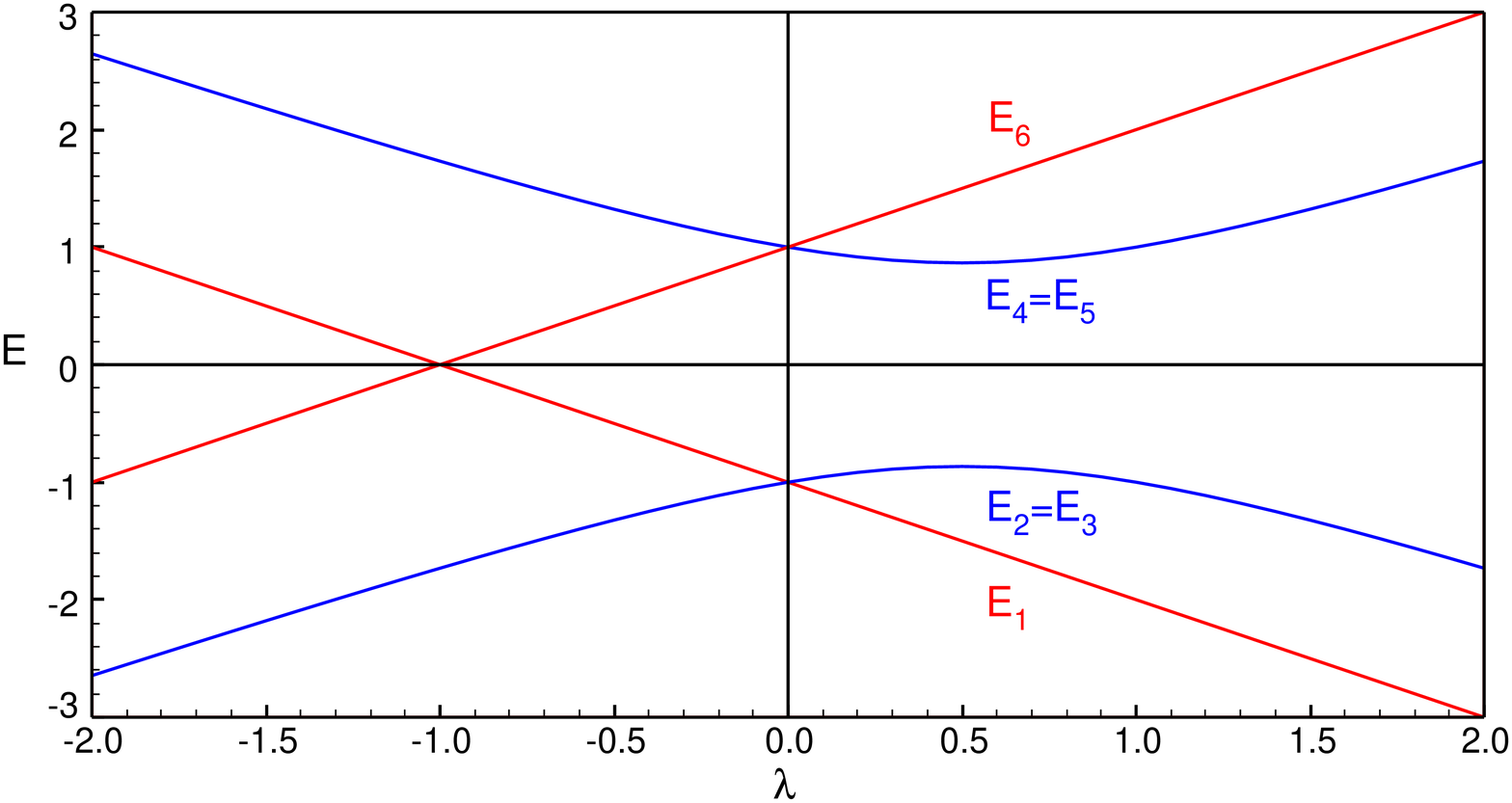}
\end{center}
\caption{Eigenvalues of the matrix (\ref{eq:H}) }
\label{Fig:Enes}
\end{figure}

Figure~\ref{Fig:Enes} shows the eigenvalues of the matrix (\ref{eq:H}) for a
range of $\lambda $ values.We appreciate the crossing between $E_{1}$ and $%
E_{6}$ at $\lambda =-1$, the crossing between the degenerate pair $\left(
E_{2},E_{3}\right) $ and $E_{1}$ at $\lambda =0$ and the crossing between
the degenerate pair $\left( E_{4},E_{5}\right) $ and $E_{6}$ also at $%
\lambda =0$. In addition to it, this figure also shows an avoided crossing
between the two degenerate pairs $\left( E_{2},E_{3}\right) $ and $\left(
E_{4},E_{5}\right) $. It is well known that only levels of different
symmetry cross, while those with the same symmetry exhibit avoided crossings
(see \cite{F14} and references therein).

In passing, we mention that the poor convergence of the perturbation series
for the resonance energy of benzene at $\lambda =1$ mentioned by Szabo and
Ostlund\cite{SO96} is due to the fact that its radius of convergence is
determined by the pair of branch points at $\lambda _{EP}^{\pm }$ and,
consequently, given by $\left| \lambda _{EP}^{\pm }\right| =1$. In other
words, $\lambda =1$ is located on the boundary of the disk of convergence.

\section{Symmetry and degeneracy}

\label{sec:symmetry}

It is well known that degeneracy can be predicted beforehand. Typically,
degeneracy is caused by the symmetry of the Hamiltonian operator. In the
present case, we expect the existence of orthogonal matrices $\mathbf{U}_{j}$%
, $j=1,2,\ldots ,N$, that leave the matrix $\mathbf{H}$ invariant; that is
to say: $\mathbf{U}_{j}^{t}\mathbf{HU}_{j}=\mathbf{H}$, where $t$ stands for
transpose and $\mathbf{U}_{j}^{t}=\mathbf{U}_{j}^{-1}$. We can rewrite this
invariance expression in terms of commutators: $\left[ \mathbf{H},\mathbf{U}%
_{j}\right] =\mathbf{HU}_{j}-\mathbf{U}_{j}\mathbf{H}=\mathbf{0}$, where $%
\mathbf{0}$ is the $6\times 6$ zero matrix.

\begin{figure}[tbp]
\begin{center}
\includegraphics[width=9cm]{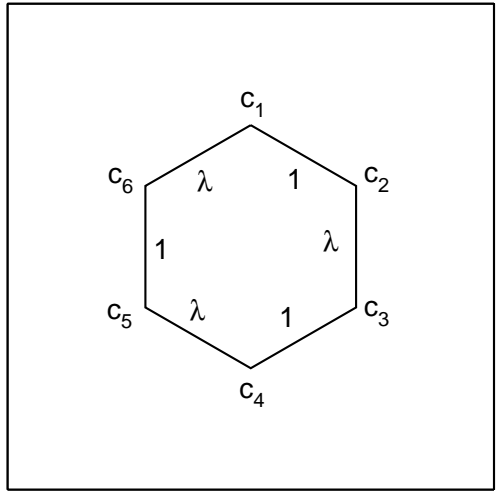}
\end{center}
\caption{Graphical representation of the matrix (\ref{eq:H}) }
\label{Fig:H6}
\end{figure}

In order to construct the orthogonal matrices just mentioned we resort to
the graphical representation of the matrix (\ref{eq:H}) shown in Figure~\ref
{Fig:H6}. The hexagon in this figure is regular when $\lambda =1$ and
irregular otherwise. Note that any rotation of $2\pi /3$ about an axis
perpendicular to the center of the figure leaves it invariant. From the
effect of this rotation: $\left[ c_{1},c_{2},c_{3},c_{4},c_{5},c_{6}\right]
\rightarrow \left[ c_{5},c_{6},c_{1},c_{2},c_{3},c_{4}\right] $, we obtain
the orthogonal matrix
\begin{equation}
\mathbf{U}_{1}=\left(
\begin{array}{llllll}
0 & 0 & 0 & 0 & 1 & 0 \\
0 & 0 & 0 & 0 & 0 & 1 \\
1 & 0 & 0 & 0 & 0 & 0 \\
0 & 1 & 0 & 0 & 0 & 0 \\
0 & 0 & 1 & 0 & 0 & 0 \\
0 & 0 & 0 & 1 & 0 & 0
\end{array}
\right) .  \label{eq:U_1}
\end{equation}
Analogously, a rotation of $4\pi /3$ about the same axis, $\left[
c_{1},c_{2},c_{3},c_{4},c_{5},c_{6}\right] \rightarrow \left[
c_{3},c_{4},c_{5},c_{6},c_{1},c_{2}\right] $, leaves the figure invariant
and is produced by the matrix
\begin{equation}
\mathbf{U}_{2}=\left(
\begin{array}{llllll}
0 & 0 & 1 & 0 & 0 & 0 \\
0 & 0 & 0 & 1 & 0 & 0 \\
0 & 0 & 0 & 0 & 1 & 0 \\
0 & 1 & 0 & 0 & 0 & 1 \\
1 & 0 & 0 & 0 & 0 & 0 \\
0 & 1 & 0 & 0 & 0 & 0
\end{array}
\right) .  \label{eq:U_2}
\end{equation}
The matrices $\mathbf{U}_{1}$ and $\mathbf{U}_{2}=\mathbf{U}_{1}^{2}$ are
representations of the point-group operations commonly called $C_{3}$ and $%
C_{3}^{2}$, respectively\cite{T64,C90}.

Figure~\ref{Fig:H6} also shows the existence of three reflection planes
perpendicular to the plane of the hexagon across the middle of opposite
sides (commonly called $\sigma _{v}$\cite{T64,C90}). The reflection $\left[
c_{1},c_{2},c_{3},c_{4},c_{5},c_{6}\right] \rightarrow \left[
c_{2},c_{1},c_{6},c_{5},c_{4},c_{3}\right] $ is produced by the matrix
\begin{equation}
\mathbf{U}_{3}=\left(
\begin{array}{llllll}
0 & 1 & 0 & 0 & 0 & 0 \\
1 & 0 & 0 & 0 & 0 & 0 \\
0 & 0 & 0 & 0 & 0 & 1 \\
0 & 0 & 0 & 0 & 1 & 0 \\
0 & 0 & 0 & 1 & 0 & 0 \\
0 & 0 & 1 & 0 & 0 & 0
\end{array}
\right) ,  \label{eq:U_3}
\end{equation}
while $\left[ c_{1},c_{2},c_{3},c_{4},c_{5},c_{6}\right] \rightarrow \left[
c_{6},c_{5},c_{4},c_{3},c_{2},c_{1}\right] $ leads to
\begin{equation}
\mathbf{U}_{4}=\left(
\begin{array}{llllll}
0 & 0 & 0 & 0 & 0 & 1 \\
0 & 0 & 0 & 0 & 1 & 0 \\
0 & 0 & 0 & 1 & 0 & 0 \\
0 & 0 & 1 & 0 & 0 & 0 \\
0 & 1 & 0 & 0 & 0 & 0 \\
1 & 0 & 0 & 0 & 0 & 0
\end{array}
\right) ,  \label{eq:U_4}
\end{equation}
and $\left[ c_{1},c_{2},c_{3},c_{4},c_{5},c_{6}\right] \rightarrow \left[
c_{4},c_{3},c_{2},c_{1},c_{6},c_{5}\right] $ is given by
\begin{equation}
\mathbf{U}_{5}=\left(
\begin{array}{llllll}
0 & 0 & 0 & 1 & 0 & 0 \\
0 & 0 & 1 & 0 & 0 & 0 \\
0 & 1 & 0 & 0 & 0 & 0 \\
1 & 0 & 0 & 0 & 0 & 0 \\
0 & 0 & 0 & 0 & 0 & 1 \\
0 & 0 & 0 & 0 & 1 & 0
\end{array}
\right) .  \label{eq:U_5}
\end{equation}

The set of matrices $G_{6}=\left\{ \mathbf{I},\mathbf{U}_{1},\mathbf{U}_{2},%
\mathbf{U}_{3},\mathbf{U}_{4},\mathbf{U}_{5}\right\} $ is a group isomorphic
to $C_{3v}$\cite{T64,C90}. If instead of the three reflection planes we
consider rotation axes on the plane of the figure across the middle of
opposite the hexagon sides, we appreciate that rotations of $\pi /2$ about
them also leave the figure invariant (they are commonly called rotation axes
$C_{2}$\cite{T64,C90}). In such a case the group $G_{6}$ is isomorphic to $%
D_{3}$\cite{T64,C90}. Both groups exhibit irreducible representations $A_{1}$%
, $A_{2}$ and $E$, the latter of dimension $2$ that explains the two-fold
degeneracy of the eigenvalues of the matrix (\ref{eq:H}) for all values of $%
\lambda $. It is clear that we can predict a vanishing discriminant from the
symmetry of the problem.

By means of projection operators\cite{T64,C90}, constructed
straightforwardly from the matrices $\mathbf{U}_{j}$, we can easily
determine the symmetry of the eigenvectors $\mathbf{v}_{j}$, $\mathbf{H}%
(\lambda )\mathbf{v}_{j}=E_{j}\mathbf{v}_{j}$, $j=1,2,\ldots ,6$, of the
matrix (\ref{eq:H}). It is not difficult to verify that $\mathbf{v}_{1}$ and
$\mathbf{v}_{6}$ are bases for the irreducible representations $A_{2}$ and $%
A_{1}$, respectively. The pairs of eigenvectors $\left( \mathbf{v}_{2},%
\mathbf{v}_{3}\right) $ and $\left( \mathbf{v}_{4},\mathbf{v}_{5}\right) $
are bases for the two-fold degenerate irreducible representation $E$.

At $\lambda =-1$ the eigenvalues $E_{1}$ and $E_{6}$ also become degenerate.
This crossing is predicted by the discriminant (\ref{eq:Disc_E(q)}) as
discussed above and is supposed to lead to an accidental degeneracy because
it does not appear to be caused by an additional symmetry based on
orthogonal matrices like those discussed above.

The highest symmetry is expected when $\lambda =1$ because the matrix $%
\mathbf{H}(1)$ is invariant under a rotation of $\pi /3$ about an axis
perpendicular to the center of the \textit{regular} hexagon in Figure~\ref
{Fig:H6}. This axis, commonly called $C_{6}$\cite{T64,C90} leads to two
additional matrix representations of $C_{6}$ and $C_{6}^{3}$ (previously we
had $C_{6}^{2}=C_{3}\rightarrow \mathbf{U}_{1}$, $C_{6}^{4}=C_{3}^{2}%
\rightarrow \mathbf{U}_{2}$). In addition to the point-group elements just
discussed we should add reflection planes $\sigma _{d}$ through opposite
vertices of the regular hexagon. The resulting group of $12$ elements is
isomorphic to $C_{6v}$. Alternatively, we may consider three axis $C_{2}$
through opposite vertices in which case the group results to be $D_{6}$\cite
{T64,C90}. In both cases the irreducible representations are $A_{1}$, $A_{2}$%
, $B_{1}$, $B_{2}$, $E_{1}$ and $E_{2}$ that also predict two-fold
degeneracy in agreement with the results in Figure~\ref{Fig:Enes}.
We do not discuss this particular case in detail because it does
not add anything relevant to what was said above.

\section{Conclusions}

\label{sec:conclusions}

Throughout this paper we have shown that the straightforward application of
the discriminant to a quantum-mechanical problem may yield a trivial useless
result if there is degeneracy for all values of the model parameter. The
cause of the failure is the symmetry of the problem which can be determined
beforehand by means of suitable tools based on group theory\cite{T64,C90}.
More precisely, one expects to face this difficulty if the symmetry of the
model Hamiltonian does not change with the variation of the model parameter.
Based on the titles of the papers by Bhattacharya and Raman\cite{BR06} and
Bhattacharya\cite{B07} we may say: \textit{Be careful if you do not look at
the spectrum}.

\appendix

\numberwithin{equation}{section}

\section{Resultant and discriminant}

\label{sec:appendix}

In this appendix we summarize some well known properties of the resultant of
two polynomials and the discriminant of a polynomial and introduce the
notation used throughout. We simply resort to what is shown in WikipediA\cite
{Wiki} and was implemented in Derive by Joseph B\"{o}hm\cite{FB07}.

The resultant of two polynomials
\begin{equation}
A(x)=\sum_{j=0}^{d}a_{j}x^{d-j},\;B(x)=\sum_{j=0}^{e}b_{j}x^{e-j},
\label{eq:A(x),B(x)}
\end{equation}
is given by the determinant
\begin{equation}
\mathrm{Res}_{x}\left( A,B\right) =\left|
\begin{array}{llllllll}
a_{0} & 0 & \cdots  & 0 & b_{0} & 0 & \cdots  & 0 \\
a_{1} & a_{0} & \cdots  & 0 & b_{1} & b_{0} & \cdots  & 0 \\
a_{2} & a_{1} & \ddots  & 0 & b_{2} & b_{1} & \ddots  & 0 \\
\vdots  & \vdots  & \ddots  & a_{0} & \vdots  & \vdots  & \ddots  & b_{0} \\
a_{d} & a_{d-1} & \cdots  & \vdots  & b_{e} & b_{e-1} & \cdots  & \vdots  \\
0 & a_{d} & \ddots  & \vdots  & 0 & b_{e} & \ddots  & \vdots  \\
\vdots  & \vdots  & \ddots  & a_{d-1} & \vdots  & \vdots  & \ddots  & b_{e-1}
\\
0 & 0 & \cdots  & a_{d} & 0 & 0 & \cdots  & b_{e}
\end{array}
\right| ,  \label{eq:Res_x(A,B)}
\end{equation}
and is related to the roots of $\lambda _{i}$, $i=1,2,\ldots ,d$, of $A(x)$
and $\mu _{j}$, $j=1,2,\ldots ,e$, of $B(x)$ by
\begin{equation}
\mathrm{Res}_{x}\left( A,B\right)
=a_{0}^{e}b_{0}^{d}\prod_{i=1}^{d}\prod_{j=1}^{e}\left( \lambda
_{i}-\mu _{j}\right) .  \label{eq:Res_x(A,B)_roots}
\end{equation}
The discriminant of $A(x)$ is given by
\begin{equation}
Disc_{x}(A)=\frac{(-1)^{d(d-1)/2}}{a_{0}}\mathrm{Res}_{x}\left(
A,A^{\prime }\right)
=a_{0}^{2d-2}\prod_{i=1}^{d-1}\prod_{j=i+1}^{d}\left( \lambda
_{i}-\lambda _{j}\right) ^{2}.  \label{eq:Disc_x(A)}
\end{equation}
Since $Disc_{x}(A)=0$ when at least two roots of $A(x)$ are equal
the discriminant of the characteristic polynomial of a
parameter-dependent matrix is suitable for the determination of
the crossings and avoided crossings of its eigenvalues.

\end{document}